\documentclass[%
 aip,
 amsmath,amssymb,
 reprint,%
]{revtex4-1}

\usepackage{graphicx}
\usepackage{dcolumn}
\usepackage{bm}
\usepackage{subfigure}

\begin{document}

\title{Evolutionary algorithm based configuration interaction approach}

\author{Rahul Chakraborty}

\author{Debashree Ghosh}
 \homepage{http://academic.ncl.res.in/debashree.ghosh}
\affiliation{%
Physical and Materials Chemistry Division, CSIR-National Chemical Laboratory, Pune 411008, India
}%
\email{debashree.ghosh@gmail.com}
\date{\today}

\begin{abstract}
A stochastic configuration interaction method based on evolutionary algorithm is designed
as an affordable approximation to full configuration interaction (FCI).
The algorithm comprises of initiation, propagation and termination steps, where the propagation
step is performed with cloning, mutation and cross-over, taking inspiration from
genetic algorithm.
We have tested its accuracy in 1D Hubbard problem and a molecular system (symmetric bond breaking
of water molecule). 
We have tested two different fitness functions based on energy of 
the determinants and the CI coefficients of determinants. We find that the absolute value of CI 
coefficients is a more suitable fitness function when combined
with a fixed selection scheme.
\end{abstract}

\maketitle


Majority of the electronic structure methods that have been developed and
used over the last few decades starts with the independent orbital
approximation, i.e. the assumption that a single Slater determinant
is a qualitatively correct starting point for a calculation.
This qualitatively correct reference is typically corrected for
dynamic correlation with post Hartree Fock (HF) methods such as
Moller-Plesset perturbation theory (MP2) or coupled cluster singles and doubles
(CCSD). However,
the assumption of a single Slater determinant as a reference 
is not qualitatively correct, especially in situations where there
are significant orbital degeneracies or near-degeneracies, e.g.,
bond breaking or di- and tri-radicals.
Such systems are referred to as strongly correlated systems and 
the electronic correlation in these systems are referred to as 
static correlation, as opposed to dynamic correlation. It is
important to note that we are not differentiating between true
correlation due to orbital degeneracies and that required to
treat proper spin symmetry (non-dynamic and static 
correlations).\cite{Reiher:JCTC:2013}

Full configuration interaction (FCI) is the most rigorous 
method to treat correlation, both static and dynamic. However,
FCI involves exact diagonalization of the full Hamiltonian in its Hilbert space
and is therefore, not affordable for reasonable system sizes and
basis sets.\cite{Knowles:FCI} Therefore, approximate methods such as CASSCF\cite{Roos:CASSCF} and RASSCF\cite{Roos:RASSCF}
etc have been developed where only a sub-set of the orbital space
is treated exactly to reduce the computational cost. But these methods
also involve an exact diagonalization, albeit over a smaller sub-space.
On the other hand, density matrix renormalization group 
(DMRG)\cite{White:DMRG,Chan:DMRG,Ghosh:DMRGSCF,Zgid:DMRGSCF,Reiher:DMRG} have
been developed to circumvent the exact diagonalization problem 
and therefore, the associated exponential scaling. While 
DMRG has been remarkably successful in the case of pseudo-linear 
systems, more general 2D and 3D systems are complicated due to problems in orbital 
ordering.\cite{Rissler:orbitalordering,Reiher:orbitalordering}
However, there have been developments towards using tensor networks to
alleviate this problem.\cite{Chan:CPS,Verstrate:Tensor}
Antisymmetrized geminal power (AGP) wavefunctions have also been a powerful technique 
for strongly correlated systems.\cite{Ortiz:geminal,Ayers:geminal}

A completely different approach is taken by stochastic algorithms such as 
quantum Monte Carlo (QMC).\cite{Umrigar:QMC:1,Umrigar:QMC:2,Neuscamman:QMC,Neuscamman:QMC:2} 
The approach uses Monte Carlo or Metropolis algorithm to propagate the 
walkers into the important part of the Hilbert space. Several flavors of QMC have been
developed, most notable of which is the FCI-QMC by Alavi and 
co-workers.\cite{Alavi:FCIQMC} There have also been other stochastic approaches
such as stochastic coupled cluster method.\cite{Thom:stochasticCC}

Evolutionary and genetic algorithm provides an alternative to 
these stochastic approaches. It has similarities to Monte Carlo Configuration
Interaction (MCCI)\cite{Greer:MCCI} and adaptive configuration interaction (ACI)\cite{Evangelista:ACI}
methods. Depending on the judicial use of 
fitness function in the evolutionary algorithm and generation of new populations, it can 
lead to extremely fast convergence to the correct solution. 
In this work, we adapt genetic algorithm (GA)
for its use in electronic structure problem. To probe the 
efficiency of the method, we have tested it for 2 systems - 1D
Hubbard model and molecular system such as
symmetric bond breaking in H$_2$O.


\begin{figure}[!htb]
\includegraphics[width=0.5\textwidth]{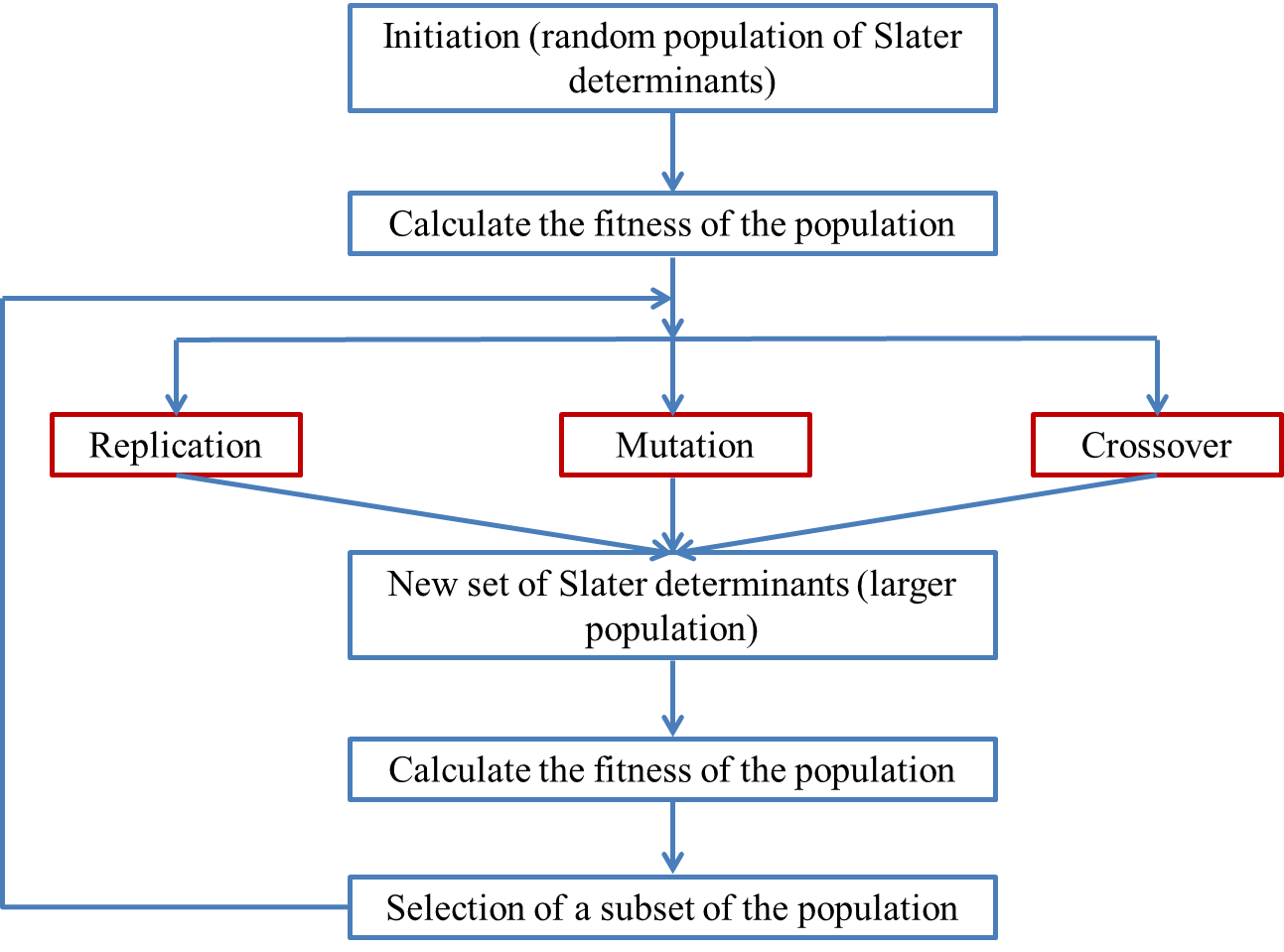}
\caption{The schematics of the algorithm used is shown. It consists of the following steps : (i) 
Initiation; (ii) Expansion of basis by replication, mutation and crossover;
(iii) Selection of the new population using the fitness function.}
\label{fig:schematics}
\end{figure}

The fermionic wavefunction of any correlated system can be written as
\begin{equation}
|\Psi\rangle = \displaystyle\sum_i c_i |D_i\rangle,
\end{equation}
where $D_i$ is the $i^{\mathrm{th}}$ Slater determinant,
\begin{equation}
|D_i\rangle = |D_{i_1,i_2...i_N}\rangle = \frac{1}{\sqrt[]{N!}}\begin{bmatrix}
\phi_{i_1}(1) & \phi_{i_1}(2) & .... & \phi_{i_1}(N) \\
\phi_{i_2}(1) & \phi_{i_2}(2) & .... & \phi_{i_2}(N) \\
. & . & . & . \\
\phi_{i_N}(1) & \phi_{i_N}(2) & .... & \phi_{i_N}(N) \\
\end{bmatrix}
\end{equation}
where the $\phi_{i_n}(m)$ denotes the occupation of electron $m$ in the orbital $i_n$.
The Hilbert space grows exponentially and therefore, the exact solution for
this FCI problem is possible for only small systems in small basis sets.

One can however, envisage solving this problem in a stochastic manner such that
the important parts of the Hilbert space (i.e.
Slater determinants with large coefficients $c_i$) are sampled in an intelligent manner.
Metropolis algorithm or Monte Carlo is one such method for stochastic sampling of the 
Hilbert space. An alternative approach is to use evolutionary algorithms.
In our genetic algorithm configuration interaction (GACI), the important steps are :
(i) Initiation; (ii) Propagation by (a) cloning, (b) mutation and (c) cross-over;
(iii) Evaluation of the fitness of the Slater determinant; and (iv) Selection, i.e.
retaining the fittest Slater determinants for the next generation population.
Using these four steps iteratively, the population of the Slater determinants
are improved and the lowest energy population is retained as the variationally
approximated wavefunction (schematics are given in Fig. \ref{fig:schematics}).
The details of each steps of the algorithm are :

\textbf{Initiation : } A GACI calculation is initiated by a random set of Slater determinants. In order 
to improve convergence the random set includes the 
determinant where the electrons (fermions)
reside in the lowest energy orbitals. In case of the molecular Hamiltonian, this
denotes the Hartree Fock wavefunction.

\begin{figure}[!htb]
\subfigure[Mutation]{
\includegraphics[width=0.28\textwidth]{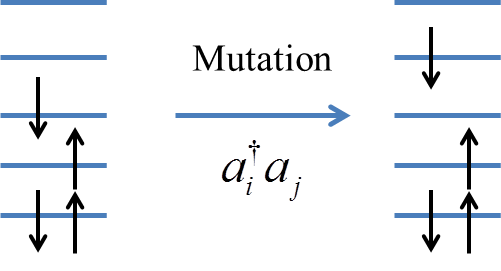}
}
\subfigure[Crossover]{
\includegraphics[width=0.5\textwidth]{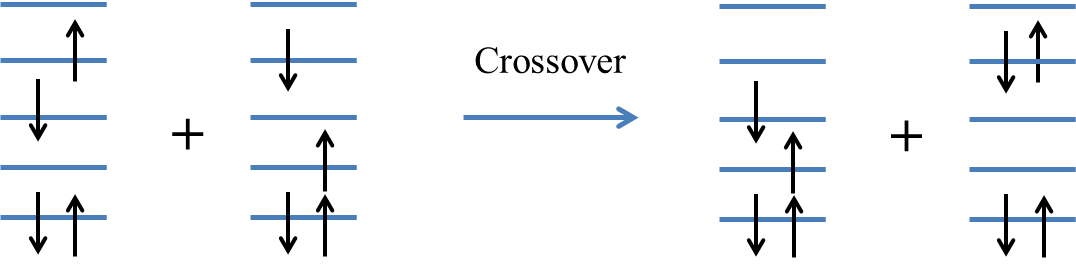}
}
\caption{The action of mutation and crossover functions on Slater determinants.}
\label{fig:mut_cross}
\end{figure}

\textbf{Propagation : } To create new generations, Slater determinants are added to the old population (set of
Slater determinants included in the previous iteration). This is done using three
different approaches with specific probabilities. The probabilities can be tuned
to achieve faster convergence. Pictorially, the different propagator operations are 
given in Fig. \ref{fig:mut_cross}.

\textbf{(i) Replication/Cloning - } The Slater determinants $|D_{i,old}\rangle$ 
in the previous iteration are always retained/copied
in the new generation.

\textbf{(ii) Mutation - } The Slater determinants are mutated with a probability $p_1$. Mutation of a
Slater determinant $|D_{i,old}\rangle$ is defined as the action of a CI singles
operator on $|D_{i,old}\rangle$, i.e.,
\begin{equation}
|D_{i}^{new}\rangle = a^\dagger_{i_k}a_{i_l}|D_{i}^{old}\rangle,
\end{equation}
such that $i_l$ is an occupied orbital, and $i_k$ is an unoccupied orbital in the old Slater determinant.

\textbf{(iii) Crossover - } The probability of a crossover function is given by, $(p_2 = 1 - p_1)$. The crossover 
function swaps the $\alpha$ and $\beta$ spin parts of the Slater determinant. 
It should be noted at this point that the GACI implemented deals only with
singlet states (i.e., where the number of $\alpha$ and $\beta$ electrons are identical)
due to the complexity of the crossover function. This issue will be
dealt with in more details in later work and generalized for all spin states.
Thus, if we expand two old generation Slater determinant as,
\begin{eqnarray}
|D_{1}^{old}\rangle &=& |D_{i,\alpha}\rangle |D_{j,\beta}\rangle \nonumber \\ 
|D_{2}^{old}\rangle &=& |D_{k,\alpha}\rangle |D_{l,\beta}\rangle 
\end{eqnarray}
where $|D_{i,\alpha}\rangle$ and $|D_{j,\beta}\rangle$ denotes the $\alpha$ and $\beta$
parts of the Slater determinant, then the new Slater determinant after the application
of the crossover operator is
\begin{eqnarray}
|D_{1}^{new}\rangle &=& |D_{i,\alpha}\rangle |D_{l,\beta}\rangle \nonumber \\ 
|D_{2}^{new}\rangle &=& |D_{k,\alpha}\rangle |D_{j,\beta}\rangle, 
\end{eqnarray}
where superscript $old$ and $new$ denotes the Slater determinant in the old and new
generations respectively. Crossover, therefore, offers a method of substantial changes
in the Slater determinant.
In the rest of the discussion, we have used $p_1 = p_2 = 0.5$. 

\textbf{Fitness function : } The proper definition of the fitness function is crucial to the efficiency of
any evolutionary algorithm. In order to estimate the efficacy of different fitness 
functions, we have implemented and tested two different forms of the fitness function.

\textbf{Energy of Slater determinant : } Each Slater determinant created by the various operations in the propagation
step has a differential occupation of orbitals. Thus, the simplest measure
of its importance in the Hilbert space is the energy of that Slater determinant. 
The energy dependent fitness function is defined as,
\begin{equation}
f(i) = \langle D_i | \hat{H} | D_i \rangle,
\end{equation}
for the Slater determinant $D_i$.

\textbf{Absolute value of CI coefficient : } In our algorithm, 
each propagation step expands the basis of Slater determinants and 
selection again brings down the total number of determinants to the required threshold population.
One can diagonalize the Hamiltonian in the expanded basis ( or the $S^{-1}H$ in the
most general case, where $S$ is the overlap between the Slater determinant basis).
The eigenvector corresponding to the lowest eigenvalue in the expanded
basis $|\tilde{\Psi}\rangle$ can be written as,
\begin{equation}
|\tilde{\Psi}\rangle = \displaystyle\sum_i c_i |D_i\rangle.
\end{equation}
The fitness function corresponding to each Slater determinant $|D_i\rangle$ is defined as,
\begin{equation}
f(i) = |c_i|,
\end{equation}
i.e. the absolute value of the CI coefficient corresponding to that Slater determinant is
used as a measure of the importance or fitness of that Slater determinant.


\textbf{Selection : } The final step of an evolutionary algorithm is a selection or death step, such
that based on a certain probability the fitter Slater determinant survives
and therefore, the next generation or population that is created is on an average
better than the previous one, in our case lower energy. There are two procedures
for the final selection or death step. 

In one case, we choose the $n$ fittest Slater determinants in a non-random fashion. 
This is denoted as fixed selection (FS).
In the other selection procedure, we use a pair-wise 
tournament selection based on a priori calculated
fitness functions and we denote this as tournament selection (TS).
The fitness functions defined above have been used as the criterion for
terminating a Slater determinant. Thus, we randomly select two Slater determinants
and retain the one whose fitness is higher than the other. This process is
continued till the size of the population reaches the threshold.


The pilot program for GACI is written in python. 
The integrals and Hamiltonians for the molecular system are 
used from \textsc{Molpro}\cite{Molpro} quantum chemistry package.
The CASSCF, MP2 and FCI results for benchmark purposes are used from the \textsc{Molpro}.

The GACI algorithm is tested for 1D Hubbard problem and for molecular system, such as bond
breaking problem in water molecule. Both these test cases have been routinely used
to probe the efficiency of methods to treat strongly correlated systems.
\begin{figure}[!htb]
\includegraphics[width=0.5\textwidth]{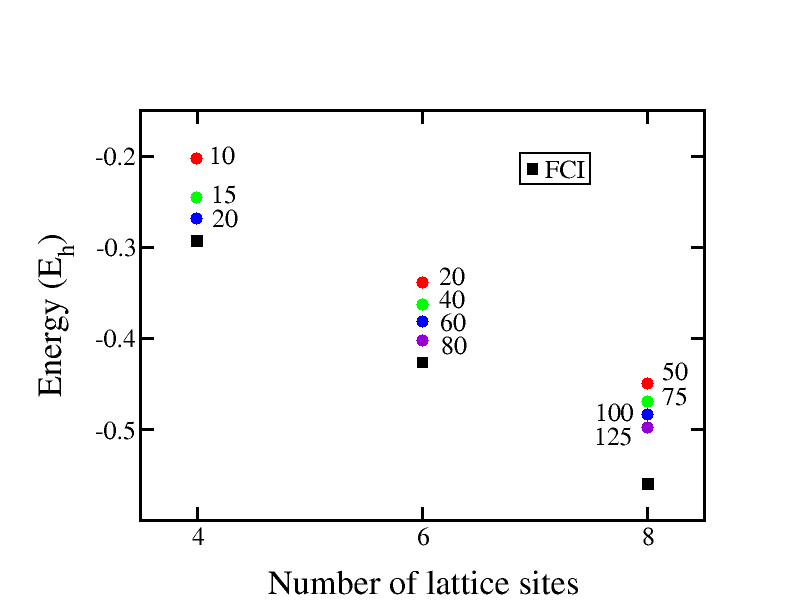}
\caption{Benchmark calculations for the comparison of GACI with FCI in the 1D Hubbard 
half filled test case for the lowest energy singlet state. 
The $U$ and $t$ chosen for each system is 10.0 and 4.0 respectively.
The number of Slater determinants retained in the GACI are given in the figure for the 4, 6 
and 8 site Hubbard systems respectively.}
\label{fig:hubbard}
\end{figure}

\textbf{1D Hubbard model system : } The Hubbard Hamiltonian is defined as,
\begin{equation}
H = -t \displaystyle\sum_{i,j,\sigma} (a^\dagger_{i\sigma}a_{j\sigma} + h.c.) + U \displaystyle\sum_j n_{j\sigma}n_{j\sigma^\prime},
\end{equation}
where $t$ is the transfer or hopping energy between sites, and $U$
is the on-site Coulomb interaction energy. The number operator is
defined as $n_{j\sigma} = a^\dagger_{i\sigma}a_{j\sigma}$.
The convergence with GACI with different population size
of Slater determinants are shown 
in Fig. \ref{fig:hubbard}, for the three system sizes that are tested (half filled). 
As expected, with the 
increase in the size of population of Slater determinants retained, the variational GACI
energy approaches the FCI limit.

\begin{figure}[!htb]
\subfigure[Potential energy surface]{
\includegraphics[width=0.5\textwidth]{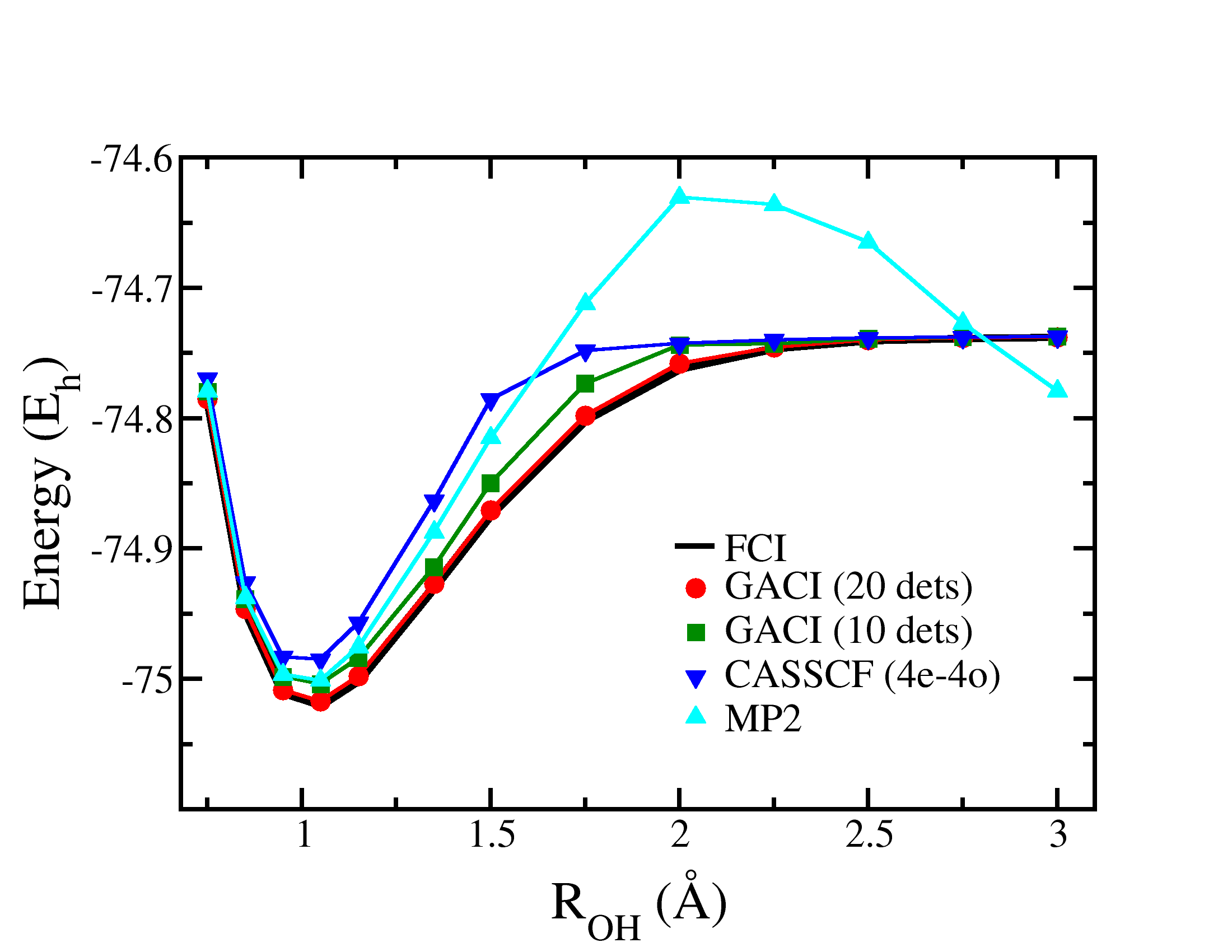}
}
\subfigure[Errors with respect to FCI]{
\includegraphics[width=0.5\textwidth]{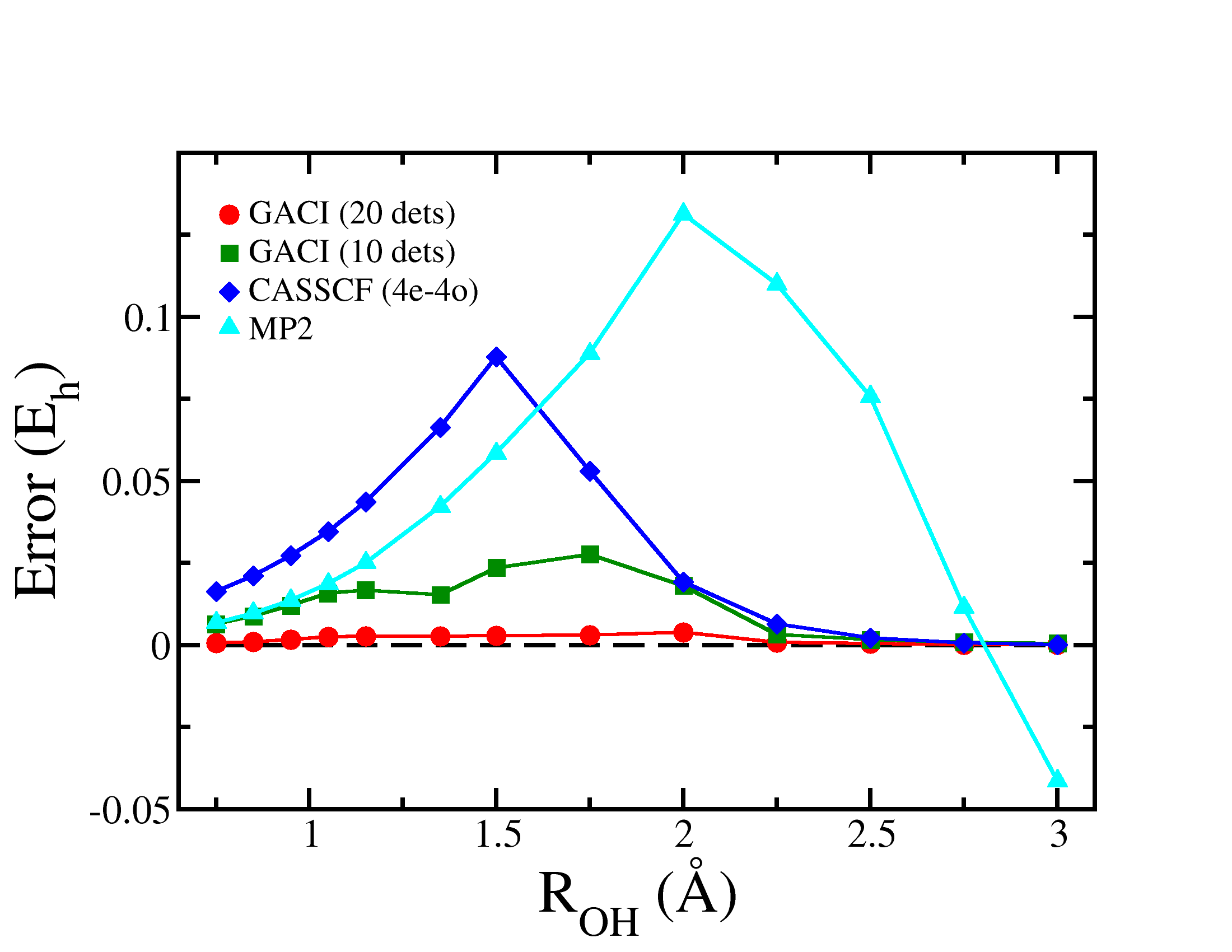}
}
\caption{Water symmetric stretching. 
(a) Potential energy surface (PES) of water molecule along symmetric bond breaking coordinate. 
Comparison of GACI with CASSCF, MP2 and FCI. (b) Error in the PES with respect to FCI 
along the bond dissociation curve.}
\label{fig:pes}
\end{figure}
\textbf{Molecular Hamiltonian : } The GACI method with fixed selection (FS) and $|c_i|$ as the fitness function
is tested for water molecule along the symmetric bond stretching mode.
The FCI and GACI energies in the STO-3G basis set are plotted
as a function of OH bond distance and compared to MP2 and CASSCF energies (Fig. \ref{fig:pes}).  
It is noticed that GACI performs much better than CASSCF and MP2 with non-parallelity errors (NPE) of 3.8 mH (20 dets) and 27.1 mH (10 dets) as compared
to 87.7 mH in CASSCF and 172.7 mH in MP2. In this case, we have calculated NPE as the difference between
the largest error and the smallest error within the limit of bond lengths for which the calculations are performed. Since the
MP2 dissociation curve turns at large bond lengths, the actual non-parallelity error of MP2 is significantly larger
than what is reported here. Comparison of the wavefunctions with GACI versus FCI shows that the overlaps 
($\langle\Psi_{\mathrm{GACI}}|\Psi_{\mathrm{FCI}}\rangle$)
are greater than 0.9 for the range of bond lengths (shown in supplementary information).

\begin{figure}[!htb]
\subfigure[Equilibrium geometry]{
\includegraphics[width=0.5\textwidth]{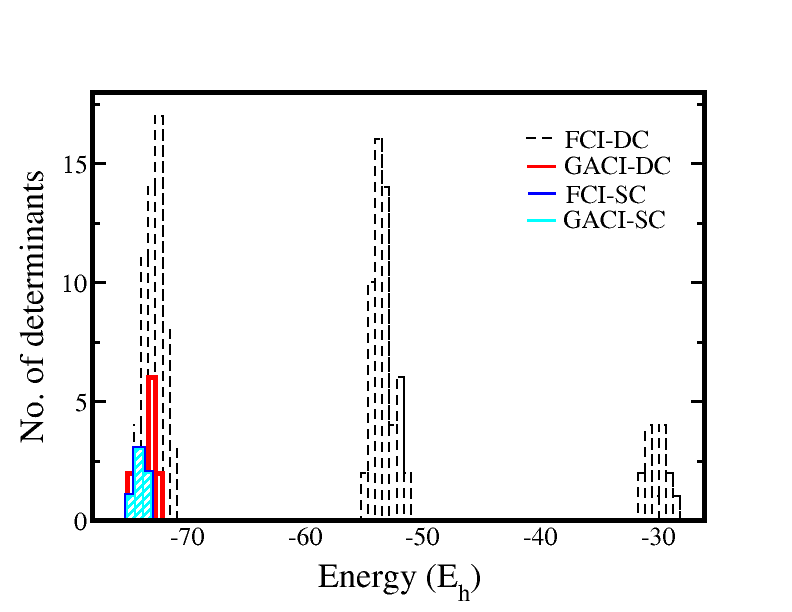}
}
\subfigure[Stretched geometry]{
\includegraphics[width=0.5\textwidth]{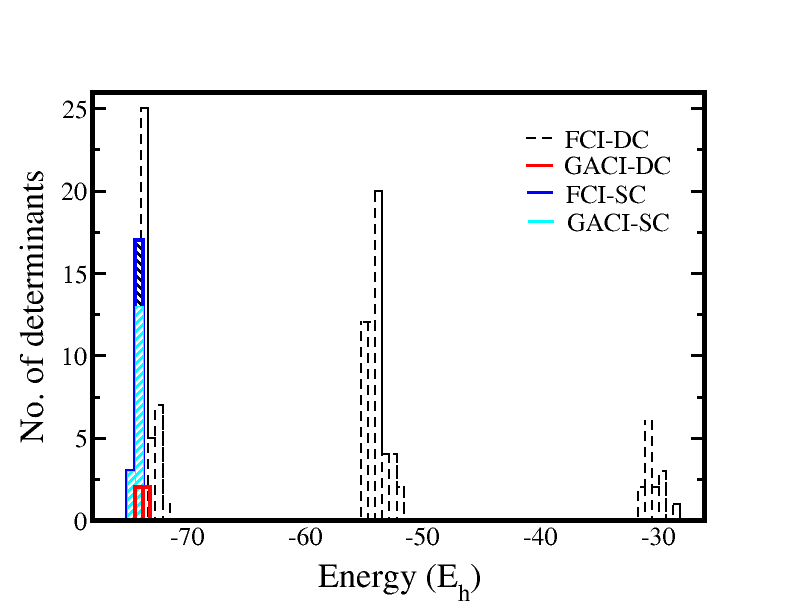}
}
\caption{SC and DC denote static and dynamic correlation respectively.
The FCI and GACI determinants with highest CI coefficients ($|c_i| \ge 0.05$) are denoted
as FCI-SC and GACI-SC. They overlap over each other showing that GACI captures all the
determinants with large coefficients. FCI-DC and GACI-DC denotes the other determinants
with smaller CI coefficients ($|c_i| < 0.05$). The GACI used is with $|c_i|$ as the fitness function and fixed selection 
criterion. (a) The histograms of numbers of determinants versus energy of determinants at the equilibrium geometry
(r$_{\mathrm{OH}}$ = 1.05 \AA\ ). (b) The histograms of numbers of determinants versus energy of determinants at 
stretched geometry (r$_{\mathrm{OH}}$ = 2.0 \AA\ ).}
\label{fig:histogram}
\end{figure}
\begin{table}[!htb]
\caption{The GACI energies (in E$_h$) with 20 determinants are shown at the equilibrium (r$_{\mathrm{OH}}$ = 1.05 \AA\ ) and one stretched
geometry (r$_{\mathrm{OH}}$ = 2.0 \AA\ ) for the two fitness functions and two selection procedures.}
\label{table:fitness}
\begin{tabular}{lcc}
\hline
Geometry & Eqm (1.05 \AA\ ) & Stretched (2.0 \AA\ )\\ 
\hline
FCI                  &  -75.019739 & -74.761988  \\
GACI (CI and FS)     &  -75.017264 & -74.758062  \\
GACI (Energy and FS) &  -75.008136 & -74.757674  \\
GACI (CI and TS)     &  -75.013062 & -74.756904  \\
GACI (Energy and TS) &  -75.009683 & -74.757453  \\
\hline
\end{tabular}
\end{table}
The comparison between the two fitness functions and selection procedures,
is shown in Table \ref{table:fitness}.
It can be seen that the absolute value of the CI coefficient is a better
fitness function and fixed selection method is a slightly better technique. 
The reason for CI coefficient being a better fitness function can be seen from Fig. \ref{fig:histogram}.
It shows the distribution of the determinants of different energies and we notice that
while the more important Slater determinants are typically with lower energy, there are 
a significant number of less important Slater determinants (with lower $|c_i|$)
that are also in the same energy range. Therefore, energy cannot be a good criterion
for the importance of a Slater determinant.

To summarize, we have developed an evolutionary algorithm based approach to
sample the important part of the Hilbert space in strongly correlated systems. It
is an alternative approach to MCCI, FCI-QMC and DMRG. It has similarities with these
methods as well as ACI. It can be combined with renormalization group approaches
to improve the subspace search in the Hilbert space. 
The convergence properties of GACI can be largely improved by changing the probabilities
of cloning, mutation and crossover. It can also be improved by the use of other fitness
functions and selection procedures. Further generation of new Slater determinants can be made
in a more efficient way using heat-bath sampling techniques.\cite{Umrigar:HBCI}
Work is in progress to rigorously test the efficiency
on these different parameters and thereby improve the convergence by improving the
creation of new generation in the important part of the Hilbert space and the acceptance ratio
of the new generations.

\vspace{0.3cm}
\noindent
\textbf{Supplementary Material}\\
See supplementary material for the most important configurations in the GACI calculations
and the overlaps between the GACI and FCI wavefunctions.

\begin{acknowledgments}
We wish to acknowledge the support from CSIR XIIth Five year plan on Multiscale
modelling for computational facilities. 
\end{acknowledgments}


\bibliography{gaci}

\end{document}